\documentclass{ceurart}

\usepackage[
  natbib = true,
    backend=biber,
    isbn=false,
    url=false,
    doi=false,
    eprint=false,
    style=numeric,
    sorting=nyt,
    sortcites = true
]{biblatex}
\AtEveryBibitem{%
	\ifentrytype{article}{
		\clearname{editor}
		\clearfield{issn}
        \clearfield{pages}
		\clearfield{urlyear}
	}{}
	\ifentrytype{inbook}{
		\clearname{editor}
		\clearfield{urlyear}
	}{}
	\ifentrytype{incollection}{
		\clearfield{isbn}
        \clearfield{pages}
		\clearfield{urlyear}
	}{}
	\ifentrytype{inproceedings}{
		\clearfield{address}
		\clearname{editor}
		\clearfield{eprint}
		\clearfield{isbn}
		\clearlist{location}
		\clearlist{publisher}
        \clearfield{volume}
        \clearfield{pages}
		\clearfield{urlyear}
		\clearfield{series}
	}{}
	\ifentrytype{proceedings}{
		\clearfield{isbn}
		\clearfield{urlyear}
	}{}
	\ifentrytype{misc}{
		\clearfield{isbn}
		\clearfield{urlyear}\clearfield{urlday}\clearfield{urlmonth} 
		\clearfield{month}
		\clearfield{eprint}
	}{}
}

\usepackage{todonotes}
\usepackage{amsmath}
\usepackage{amssymb}
\usepackage{amsthm}
\usepackage[linesnumbered,ruled,vlined]{algorithm2e}
\DontPrintSemicolon
\SetKwInOut{Input}{Input}
\SetKwInOut{Output}{Output}
\SetKwInOut{Data}{Data}
\SetAlFnt{\footnotesize}
\SetAlCapFnt{\small}
\SetAlCapNameFnt{\small}
\SetArgSty{textnormal}

\usepackage{cleveref}
\usepackage{subcaption}
\usepackage{booktabs}
\usepackage{float}
\usepackage{array}
\usepackage{rwthcolors}

\newtheorem{definition}{Definition}[section]

\newtheorem{example}{Example}[section]

\DeclareRobustCommand{\hlremove}[1]{\textcolor{blue75!50!black}{\underline{#1}}}
\DeclareRobustCommand{\hlalgo}[1]{{\setlength{\fboxsep}{0pt}\colorbox{black!20!white}{#1}}}
\DeclareRobustCommand{\hlmath}[1]{{\setlength{\fboxsep}{0pt}\colorbox{black!20!white}{\ensuremath{#1}}}}

\newcommand\Undef[0]{\textit{undef}\xspace}
\newcommand\false[0]{\textit{false}\xspace}
\newcommand\true[0]{\textit{true}\xspace}

\newcommand\vdeg[2]{\deg_{#1}(#2)}

\DeclareMathOperator{\levelOp}{level}
\newcommand\level[1]{\levelOp(#1)}

\DeclareMathOperator{\discOp}{disc}
\newcommand\disc[2]{\discOp_{#1}(#2)}
\DeclareMathOperator{\resOp}{res}
\newcommand\res[3]{\resOp_{#1}(#2,#3)}

\DeclareMathOperator{\realRootsOp}{realRoots}
\newcommand\realRoots[1]{\realRootsOp(#1)}

\DeclareMathOperator{\irootOp}{root}
\newcommand\iroot[3]{\irootOp_{#1}^{#2,#3}}

\newcommand\rationals[0]{\mathbb{Q}}
\newcommand\reals[0]{\mathbb{R}}

\newcommand\naturals[0]{\mathbb{N}}

\newcommand\posints[0]{\mathbb{N}_{>0}}

\newcommand\proj[2]{\ensuremath{#1_{#2}}}

\newcommand\pto[0]{\hookrightarrow}

\newcommand\Isymb[0]{\textnormal{\texttt{I}}}
\newcommand\Tsymb[0]{\textnormal{\texttt{T}}}

\newcommand\range[2]{\ensuremath{[#1..#2]}}
\newcommand\prange[1]{\ensuremath{[#1]}}
\newcommand\zrange[1]{\ensuremath{[0..#1]}}

\newcommand\cell[1]{C(#1)}

\newcommand{\eoe}{\ensuremath{\lrcorner}} %
\newcommand{\eod}{\ensuremath{\Box}} %

\begin{document}

\copyrightyear{2025}
\copyrightclause{Copyright for this paper by its authors.
  Use permitted under Creative Commons License Attribution 4.0
  International (CC BY 4.0).}

\conference{SC² 2025: Satisfiability Checking and Symbolic Computation}

\title{A Variant of Non-uniform Cylindrical Algebraic Decomposition for Real Quantifier Elimination}

\tnotemark[1]
\tnotetext[1]{Jasper Nalbach was supported by the Deutsche Forschungsgemeinschaft (DFG, German Research Foundation) as part of RTG 2236 \emph{UnRAVeL} and AB 461/9-1 \emph{SMT-ART}. Computations were performed with computing resources granted by RWTH Aachen University under project rwth1560.}

\author[1]{Jasper Nalbach}[orcid=0000-0002-2641-1380,email=nalbach@cs.rwth-aachen.de]
\cormark[1]
\author[1]{Erika {\'A}brah{\'a}m}[orcid=0000-0002-5647-6134,email=abraham@cs.rwth-aachen.de]

\address[1]{RWTH Aachen University, Aachen, Germany}

\cortext[1]{Corresponding author.}

\begin{abstract}
  The \emph{Cylindrical Algebraic Decomposition (CAD) method} is currently the only complete algorithm used in practice for solving real-algebraic problems. To ameliorate its doubly-exponential complexity, different \emph{exploration-guided} adaptations try to avoid some of the computations. The first such adaptation named \emph{NLSAT} was followed by \emph{Non-uniform CAD (NuCAD)} and the \emph{Cylindrical Algebraic Covering (CAlC)}. Both NLSAT and CAlC have been developed and implemented in SMT solvers for satisfiability checking, and CAlC was recently also adapted for quantifier elimination. However, NuCAD was designed for quantifier elimination only, and no complete implementation existed before this work.
In this paper, we present a novel variant of NuCAD for both real quantifier elimination and SMT solving, provide an implementation, and evaluate the method by experimentally comparing it to CAlC.
\end{abstract}


\maketitle

\section{Introduction}

\emph{Real algebra} is a highly expressive logic, whose formulas are Boolean combinations of polynomial constraints over potentially quantified real-valued variables. In this paper, we consider three related problems for real algebra: (1) deciding the satisfiability of quantifier-free formulas (i.e., solving the existential fragment), (2) deciding the truth of sentences, and (3) quantifier elimination.

The first practically feasible complete algorithm for solving all these problems was the \emph{cylindrical algebraic decomposition (CAD) method} \cite{collins1975}. Though doubly-exponential in complexity, CAD and some of its variants are till today the only complete solutions offered in  computer algebra tools like \texttt{QEPCAD B} \cite{brown1999,brown2003}, \texttt{Redlog} \cite{seidl2003}, and commercial systems like \texttt{Mathematica} \cite{strzebonski2000} or \texttt{Maple} \cite{chen2009,iwane2009}.

On another research line, \emph{exploration-guided proof generation} turned out to be extremely effective for checking the satisfiability of propositional logic formulas via SAT solvers. Extending this idea to \emph{Satisfiability Modulo Theories (SMT) solving}, the \emph{NLSAT} \cite{jovanovic2012} algorithm was the first exploration-guided CAD adaptation for satisfiability checking. Whereas CAD analyses the whole state space regarding the whole input formula, NLSAT first explores the state space through smart guesses of sample values for the (real and Boolean) variables. If a partial sample turns out not to be further extensible to a satisfying assignment, then some partial CAD computations are applied to generalize this sample to a set of samples, which are not extensible to satisfying solutions for the same reason (i.e., the same constraints in the input formula). These generalizations rely on the \emph{single cell} paradigm introduced in \cite{brown2013,brown2015} and refined in \cite{nalbach2024levelwise}.
NLSAT implementations exist in the SMT solvers \texttt{z3} \cite{demoura2008}, \texttt{yices2} \cite{dutertre2014}, and \texttt{SMT-RAT} \cite{corzilius2015}.

NLSAT inspired the development of another CAD adaptation named the \emph{Cylindrical Algebraic Covering (CAlC)} \cite{abraham2021} algorithm. It also uses exploration and single-cell generalization, but it reduces the bookkeeping effort of NLSAT (to remember all the generalizations). Though the original CAlC algorithm cannot handle Boolean structures other than conjunctions and thus this task needs to be outsourced to a SAT solver, such a (lazy SMT) CAlC implementation in \texttt{cvc5} \cite{kremer2021calcimpl} outperforms NLSAT in \texttt{z3} and \texttt{yices2}.

While having a focus on satisfiability checking, some SMT solvers can also solve quantified formulas \cite{demoura2007,bjorner2015,niemetz2021}. While most such support is incomplete for the theory of the reals, the \emph{QSMA} \cite{bonacina2023} algorithm implemented in \texttt{yicesQS}, which relies on NLSAT, is complete.
Recently, also the CAlC algorithm was extended to transfer the performance gains from the SMT world to quantifier elimination \cite{kremer2023calc,nalbach2024calc}, successfully outperforming (in terms of running time) existing SMT solvers and open source implementations supporting quantifier elimination.

Also the \emph{non-uniform cylindrical algebraic decomposition (NuCAD)} \cite{brown2015nucad} builds on the NLSAT idea to decompose $\reals^n$ similar to the CAD, but with fewer cells. Its first version was not able to reason about quantifier alternations. The work in \cite{brown2017nucad} suggests to first compute a NuCAD of $\reals^n$, and then to refine this decomposition to respect the quantifier structure of the input formula. NuCAD as introduced in these papers is incomplete in the sense that it considers only \emph{open cells}, i.e., it might not be able to solve formulas with non-strict inequalities.

\smallskip
\noindent Our paper continues the above work on the NuCAD, contributing:
\vspace*{-0.75ex}
\begin{itemize}
    \item a \emph{complete} variant of the NuCAD algorithm,
    \item a modification which considers the input's quantifier structure \emph{directly} during the computation of the decomposition,
    \item its \emph{implementation} along with a \emph{post-processing} of the NuCAD, and
    \item \emph{experimental evaluation} in particular against the CAlC algorithm, on deciding the existential fragment, deciding sentences, and quantifier elimination.
\end{itemize} 
Though our experiments indicate that CAlC solves satisfiability checking problems faster than NuCAD, NuCAD seems to be competitive on quantifier elimination. However, the available quantified benchmark sets lack sufficient diversity to make definitive conclusions on algorithm superiority.

\paragraph{Outline:}
We introduce the fundamentals of CAD and CAlC in \Cref{sec:preliminaries}, and introduce a complete version of the NuCAD algorithm in \Cref{sec:nucad}. In \Cref{sec:nucad:quantifiers}, we extend the algorithm for quantifier alternations. %
Finally, we present the experimental results in \Cref{sec:experiments} and conclude the paper in \Cref{sec:conclusion}.

\section{Preliminaries}
\label{sec:preliminaries}

Let $\naturals$, $\posints$, $\rationals$, and $\reals$ denote the natural (incl. $0$), positive integer, rational, and real numbers respectively. 
For $i,j\in\naturals$, we set $[i..j] = \{ i,\ldots,j \}$ and $[i] = [1..i]$.
For $j\in\naturals$, $s\in\reals^j$ and $k\in [j]$ we write $s_k$ for the $k$th component of $s$.
For tuples introduced as $t = (a,b,c)$, we use $t.a$, $t.b$ and $t.c$ to access their entries.

\noindent From now $n\in\posints$, $i \in \zrange{n}$,$j\in\prange{n}$, $s\in \reals^j$, $s^* \in \reals$, $R \subseteq \reals^j$, and $I \subseteq \reals$.

We define $(s, s^*)=(s_1,\ldots,s_j,s^*)$, $s \times I = \{ s \} \times I $, $\proj{s}{[i]}=(s_1,\ldots,s_{\min\{i,j\}})$, and  $\proj{R}{[i]}=\{\proj{s}{[i]}|s\in R\}$.
Let $f,g: R \to \reals$ be continuous functions, then $R \times (f,g) = \{ (r,r') \mid r \in R \wedge r' \in \reals \wedge f(r) < r' < g(r)  \}$, $R \times [f,g] = \{ (r,r') \mid r \in R \wedge r' \in \reals \wedge f(r) \leq r' \leq g(r)  \}$, analogously $R \times (f,g]$ and $R \times [f,g)$.

Throughout this paper, we assume ordered \emph{variables} $x_1 \prec \cdots \prec x_n$. Let $\rationals[x_1,\ldots,x_i]$ be the set of all \emph{polynomials} in $x_1,\ldots,x_i$ with rational coefficients. Let $p\in \rationals[x_1,\ldots,x_i]$. The \emph{main variable} of $p$ is the highest ordered variable occurring in it, and its \emph{level} $\level{p}$ is the index of the main variable.
The \emph{degree} of $p$ in $x_j$ is denoted by $\vdeg{x_j}{p}$.
For a univariate $p \in \rationals[x_j]$, its \emph{real roots} (in $x_j$) build the set $\realRoots{p}$.

Let $p(s)$ result from $p$ by substituting the values $s_k$ for $x_k$, for $k\in \prange{i}$. 
We call $p$ \emph{sign-invariant} on $R'\subseteq\reals^i$ if $p(s')$ has the same sign for all $s'\in R'$. 
The polynomial $p$ is \emph{nullified over $s$} if $p(s) \equiv 0$, where $\equiv$ denotes semantic equivalence.
We further define $p[s]$ to be $p(s)$ for $\level{p}\leq j$, and $p$ otherwise.

A (polynomial) \emph{constraint} has the form $p \sim 0$ with $\sim \in \{=,\leq,\geq,\neq,<,>\}$. Notations for polynomials are transferred to constraints where meaningful, for example $(p\sim 0)(s)$ is $p(s)\sim 0$, and $(p\sim 0)[s]$ is $p[s]\sim 0$.

Real-algebraic \emph{formulae} $\varphi$ are potentially quantified Boolean combinations of polynomial constraints. Again, we inherit notations from constraints, e.g. we get $\varphi(s)$ ($\varphi[s]$) from $\varphi$ through replacing each constraint $c$ by $c(s)$ ($c[s]$).
A formula $\varphi$ is \emph{truth-invariant} on $R' \subseteq \reals^i$ if $\varphi(s')=\varphi(s'')$ for all $s',s''\in R'$.
A formula $\varphi$ is in \emph{prenex normal form} if it is of the form  $Q_{k+1} x_{k+1}\ldots Q_n x_{n}.\; \bar{\varphi}$ for some $k\in\naturals$, consisting of a \emph{prefix} of quantifiers and a quantifier-free formula $\bar{\varphi}$ called the \emph{matrix}; the variables  $x_{1},\ldots,x_{k}$ are called \emph{free} in $\varphi$ (also called \emph{parameters}).
The task of \emph{quantifier elimination} is, given a formula $\varphi$ that may contain quantifiers, to compute a quantifier-free formula $\psi$ such that $\psi \equiv \varphi$.

We will further make use of the following concept:

\begin{definition}[Implicant \cite{kremer2023calc,nalbach2024calc}] \label{def:implicant}
	Let $j \in \naturals$, $s \in \reals^j$, and $\varphi$ be a formula where (at least) the variables $x_1,\ldots,x_j$ are free.

    The quantifier-free formula $\psi$ is an \emph{implicant of $\varphi$ with respect to $s$} if all constraints $p\sim 0$ in $\psi$ have $\level{p}\leq j$ and are contained in $\varphi$, %
    $\psi[s] \equiv \true$, and either $\psi \Rightarrow \varphi$ or $\psi \Rightarrow \neg\varphi$.
    \hfill\eod
\end{definition}

\begin{example}
	As a univariate example with $j=1$, for $\varphi = x_1^2>0 \wedge (x_1<2 \vee x_1>4)$ we have $\varphi[1] \equiv \true$, $\varphi[3] \equiv \false$, and $\varphi[0] \equiv \false$. Examples for implicants of $\varphi$ are $x_1^2>0 \wedge x_1 < 2$ w.r.t. the value $s=1(\in\reals^1)$, $\neg (x_1<2 \vee x_1>4)$ w.r.t. $3$, and both $\neg(x_1^2>0)$ and $\neg(x_1^2>0 \wedge x_1>4)$ are implicants of $\varphi$ w.r.t. $0$.

	For $\varphi' = (x_1<0 \vee x_2\leq4)\wedge (x_1>2 \vee x_2>4)$ we observe $\varphi'[1] \equiv \false$, and identify $\neg(x_1<0) \wedge \neg(x_1>2)$ to be an implicant of $\varphi'$ w.r.t. $1$.\hfill\eoe
\end{example}

\subsection{Cylindrical Algebraic Decomposition and
  Covering}

For a given formula, the \emph{cylindrical algebraic decomposition (CAD)} \cite{collins1975} method decomposes $\reals^n$ into a finite number of subsets such that each polynomial in the formula is sign-invariant on each subset. The finiteness of the decomposition allows to reason about the solution space of the formula. Furthermore, to admit quantifier elimination, these sets are arranged in a cylindrical structure.

\begin{definition}[Cylindrical Algebraic Decomposition \cite{collins1975}]
    Let $i \in \prange{n}$.
    \begin{itemize}
      \item An \emph{($i$-dimensional) cell} is a non-empty connected subset of $\reals^i$.
    \item A \emph{decomposition} of $\reals^i$ is a finite set $D_i$ of $i$-dimensional cells such that either $R=R'$ or $R\cap R'=\emptyset$ for all $R,R'\in D_i$, and $\cup_{R\in D_i} R=\reals^i$.
    \item A set $R' \subseteq \reals^i$ is \emph{semi-algebraic} if it is the solution set of a quantifier-free formula. 
        \item A decomposition $D_i$ of $\reals^i$ is \emph{cylindrical} if either $i=1$, or $i>1$ and $D_{i-1} = \{ \proj{R}{[i-1]} \mid R \in D_i \}$ is a cylindrical decomposition of $\reals^{i-1}$. %
        \item A \emph{cylindrical algebraic decomposition (CAD) of $\reals^i$} is a cylindrical decomposition $D$ of $\reals^i$ whose cells $R \in D$ are semi-algebraic.
        \item Let $P \subseteq \rationals[x_1,\ldots,x_n]$ and $D$ be a CAD of $\reals^n$. We call $D$ \emph{sign-invariant for $P$} if each $p \in P$ is sign-invariant on each cell $R \in D$.
        \item Let $R \subseteq \reals^i$ be a cell and $f,g: R \to \reals$ continuous functions with $f(s)<g(s)$ for all $s\in R$. We call $R \times \reals$ the \emph{cylinder over $R$}, $R \times [f,f]$ a \emph{section over $R$}, and $R \times (f,g)$ a \emph{sector over $R$}.\hfill\eod
    \end{itemize}
\end{definition}

\noindent Unfortunately, the number of cells in a CAD that is sign-invariant for some set of polynomials may be doubly exponential in the number of their variables. This heavy complexity is partly rooted in the \emph{projection phase} of the CAD method, in which polynomials of level $i$ are projected to generate polynomials of level $i-1$, iteratively for $i=n,\ldots,2$.
The effort for these projections depends on the degree and on the number of the polynomials, which might grow double-exponentially during this process. Thus, the costs of the projection plays a substantial role for the practical feasibility of the algorithm.

\begin{figure}[t]
    \begin{subfigure}[t]{0.49\textwidth}
        \centering
        \includegraphics[width=\textwidth]{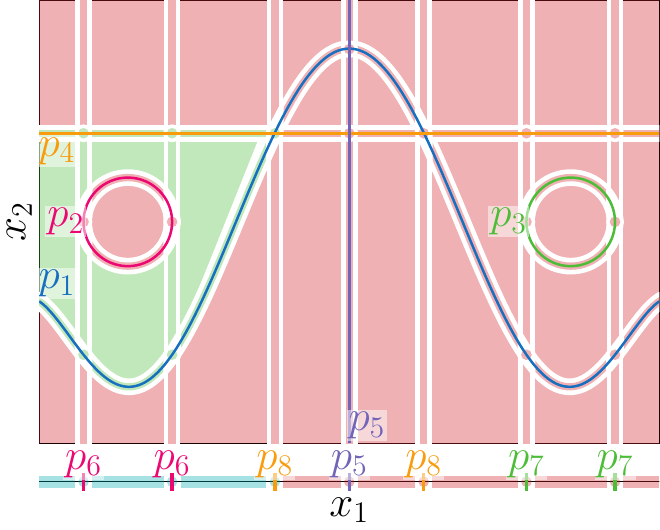}
    
        \caption{Sign-invariant CAD from \Cref{ex:cad}.}
        \label{fig:cad}
    \end{subfigure}\hfill%
    \begin{subfigure}[t]{0.49\textwidth}
        \centering
        \includegraphics[width=\textwidth]{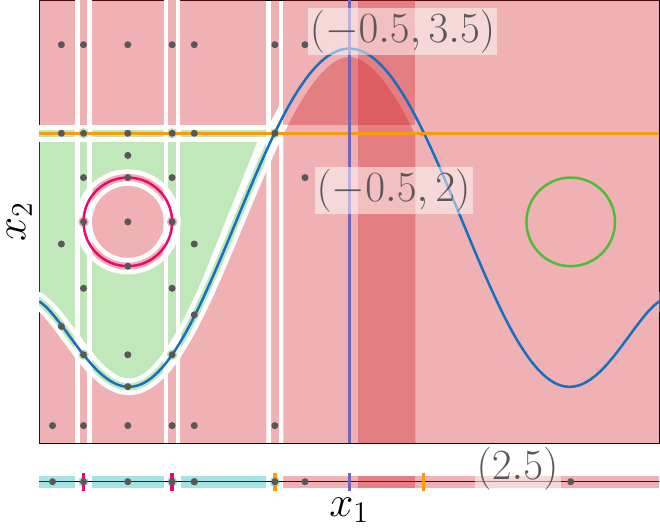}
    
        \caption{Truth-invariant CAlC from \Cref{ex:calc}, along with a sample point for each cell.}
        \label{fig:calc}
    \end{subfigure}
    \caption{In the two-dimensional coordinate system, red cells evaluate the formula 
    $\varphi$
    to $\false$, and green cells evaluate $\varphi$ to $\true$. In the one-dimensional coordinate system, red cells evaluate  $\varphi$ to $\false$, and in blue cells $\varphi$ is satisfiable.}
\end{figure}

\begin{example}\label{ex:cad}
    Consider the formula $\varphi(x_1,x_2) = p_1 \leq 0 \wedge p_2 > 0 \wedge p_3 \geq 0 \wedge p_4 \leq 0 \wedge p_5 \leq 0$ built using the following polynomials:
    \begin{align*}
      \multispan{4}{$p_1 = -0.006(x_1-2)(x_1+2)(x_1-3)(x_1+3)(x_1-4)(x_1+4)-x_2$}\\
      p_2 = (x_1+2.5)^{2}+(x_2-1.5)^{2}-0.25 & \hspace*{5ex} & p_4 &= x_2-2.5\\
      p_3 = (x_1-2.5)^{2}+(x_2-1.5)^{2}-0.25 && p_5 &= x_1
    \end{align*}
    
    \noindent The CAD method computes the projection polynomials $p_6 = \disc{x_2}{p_2}$, $p_7 = \disc{x_2}{p_3}$ and $p_8 = \res{x_2}{p_1}{p_4}$, where disc and res are operators which we do not detail here. 
    \Cref{fig:cad} illustrates the variety of these polynomials, as well as the CAD of $\reals$ and $\reals^2$.

    Further, each cell is coloured according to the truth value (green for $\true$ and red for $\false$) of $\varphi$ on that cell, which is determined by substituting a sample point (not depicted in the figure) from each cell into $\varphi$.
    
    If we aim to check whether $\exists x_1.\; \exists x_2.\; \varphi$ holds, we only need to find one cell where $\varphi$ evaluates to $\true$. To check whether $\exists x_1.\; \forall x_2.\; \varphi$ holds, we would iterate through the one-dimensional CAD (e.g. finitely many values for $x_1$) and then check whether the respective cylinder is covered by cells where $\varphi$ is $\true$.\hfill\eoe
\end{example}

A key observation is that the computed CAD is often finer than necessary for quantifier elimination. Some projection polynomials may be omitted, leading to a coarser CAD, because the real roots of the projection polynomials define the boundaries of the CAD cells. Now the idea of \emph{exploration-guided} algorithms comes into play: we can iteratively guess sample points and generalize them to truth-invariant cells, and compute a coarser decomposition (or covering) of the state space by combining the individual generalizations. Such algorithms are \emph{NuCAD} with a weaker notion of cylindricity, \emph{CAlC} \cite{abraham2021} which maintains cylindricity but relaxes decomposition to covering, and \emph{NLSAT} \cite{jovanovic2012} without an explicit cell structure but implicitly computes a something similar as CAlC (and is not directly extended for QE yet). In \Cref{sec:experiments} we will compare NuCAD against CAlC, therefore we introduce it below.

\begin{definition}[Cylindrical Algebraic Covering \cite{abraham2021}]
    Let $i \in \prange{n}$.
    \begin{itemize}
    \item A \emph{covering} of $\reals^i$ is a finite set $D_i$ of $i$-dimensional cells with $\cup_{R\in D_i} R=\reals^i$.
    \item A covering $D_i$ of $\reals^i$ is \emph{minimal} if there does not exist a covering $D'_i \subseteq D_i$ of $\reals^i$.
    \item A covering $D_i$ of $\reals^i$ is \emph{cylindrical} if either $i=1$, or $i>1$ and $D_{i-1} = \{ \proj{R}{[i-1]} \mid R \in D_i \}$ is a minimal cylindrical covering of $\reals^{i-1}$. 
    \item A \emph{cylindrical algebraic covering (CAlC) of $\reals^i$} is a cylindrical covering $D_i$ of $\reals^i$ whose cells $R \in D_i$ are semi-algebraic.
    \item Let $\varphi$ be a formula with free variables $x_1,\ldots,x_n$. A CAlC $D_n$ of $\reals^n$ is \emph{truth-invariant for $\varphi$} if $\varphi$ is truth-invariant on each cell $R \in D_n$.\hfill\eod
    \end{itemize}
\end{definition}

\begin{example}\label{ex:calc}
	A truth-invariant CAlC for the formula $\varphi$ from \Cref{ex:cad} is depicted in \Cref{fig:calc}. We emphasize that due to the cylindrical arrangement, we can reason about the formula and its quantifiers similarly as with a CAD.\hfill\eoe
\end{example}

\subsection{Single Cylindrical Cells}

A key concept for exploration-guided algorithms is the \emph{single cell construction} \cite{brown2013,brown2015,nalbach2024levelwise}, which takes a set of polynomials and a sample point as input, and computes the description of a cell that contains the sample point, such that each input polynomial is sign-invariant over the cell.
These cells have the following property, which allows to describe them explicitly:

\begin{definition}[Local Cylindricity \cite{brown2015}]
    Let $i \in \prange{n}$.
    A set $R\subseteq\reals^i$ is \emph{locally cylindrical} if either $i=1$, or $i>1$ and $R$ is a section or sector over $\proj{R}{[i-1]}$ which itself is locally cylindrical.\hfill\eod
\end{definition}

\noindent We can thus describe locally cylindrical cells by bounds on $x_1$, bounds on $x_2$ that depend on $x_1$, bounds on $x_3$ that depend on $x_1$ and $x_2$, and so on. These bounds are described as varieties of some projection polynomials, e.g. each bound is some root of a polynomial. We formalize these notions using symbolic intervals.

\begin{definition}[Indexed Root Expression, Symbolic Interval]
        An \emph{indexed root expression (of level $i$)} is a partial function $\iroot{x_i}{p}{j}: \reals^{i-1} \pto \reals$ for some $i,j\in\posints$ and $p \in \rationals[x_1,\ldots,x_i]$ with $\level{p}=i$, such that for each $s\in\reals^{i-1}$, $\iroot{x_i}{p}{j}(s)$ is the $j$-th real root of the univariate polynomial $p(s)\in\rationals[x_i]$ if it exists, and it is undefined (\Undef) otherwise.

        A \emph{symbolic interval $\Isymb$ of level $i$}  has the form $\Isymb = (l, u)$, $\Isymb = [l, u]$, $\Isymb = [l, u)$, or $\Isymb = (l, u]$ where $l$ is either an indexed root expression of level $i$ or $-\infty$, and $u$ is either an indexed root expression of level $i$ or $\infty$. The polynomials $\Isymb.l$ and $\Isymb.u$ are the \emph{defining polynomials of $\Isymb$} (if they exist).\hfill\eod
\end{definition}

We use the cross product notation also for symbolic intervals.

\SetKwFunction{FChooseInterval}{choose\_interval}
\SetKwFunction{FComputeCellProjection}{compute\_cell\_projection}
\SetKwFunction{FConstructCell}{construct\_cell}
\begin{algorithm}[b]
    \caption{\protect\FConstructCell{$P$,$s$}.}
    \label{alg:cell:construct}

    \Input{$P=P' \subseteq \rationals[x_1,\ldots,x_n]$, $s \in \reals^n$.}
    \Output{Symbolic intervals $\Isymb_1,\ldots,\Isymb_n$ of levels $1,\ldots,n$ respectively such that $\Isymb_1 \times \ldots \times \Isymb_n \subseteq \cell{P,s}$.}

    \ForEach{$i=n,\ldots,1$}{
        $\Isymb_i := $ \FChooseInterval{$P'$,$s_{[i]}$} \tcp*{\Cref{alg:cell:choose-interval}}
        $P' := $ \FComputeCellProjection{$P'$,$s_{[i]}$,$\Isymb_i$} \tcp*{\Cref{alg:cell:compute-cell-projection}}
    }
	\Return{$\Isymb_1,\ldots,\Isymb_n$}
\end{algorithm}

\begin{algorithm}[t]
    \caption{\protect\FChooseInterval{$P$,$s$}.}
    \label{alg:cell:choose-interval}

    \Input{$P \subseteq \rationals[x_1,\ldots,x_i]$, $s \in \reals^i$ for an $i \in \prange{n}$.}
    \Output{Symbolic interval $\Isymb$ of level $i$ such that $s_{[i-1]} \times \Isymb = s_{[i-1]} \times \{ r \in \reals \mid (s_{[i-1]},r) \in \cell{P,s} \}$.}

    $\Xi := \cup_{p \in P} \realRoots{p(s_{i{-}1})}$ \;
    \lIf{$\Xi=\emptyset$}{ \Return{$(-\infty,\infty)$}}
    {otherwise assume} $\Xi=\{ \xi_1,\ldots,\xi_k \} $ such that $\xi_1 < \ldots < \xi_k$ \;

    \If{$s_i \in (-\infty,\xi_1)$}{
        \Return{$(-\infty,\iroot{x_i}{p}{j})$ for $p \in P$, $j \in \posints$ such that $\xi_1 = \iroot{x_i}{p}{j}(s_{[i-1]})$}
    }
    \ElseIf{$s_i \in (\xi_\ell,\xi_{\ell+1})$}{
        \Return{$(\iroot{x_i}{p}{j},\iroot{x_i}{p'}{j'})$ for $p,p' \in P$, $j,j' \in \posints$ such that $\xi_\ell = \iroot{x_i}{p}{j}(s_{[i-1]})$ and $\xi_{\ell+1} = \iroot{x_i}{p'}{j'}(s_{[i-1]})$} 
    }
    \ElseIf{$s_i = \xi_\ell$}{
        \Return{$[\iroot{x_i}{p}{j},\iroot{x_i}{p}{j}]$ for $p \in P$, $j \in \posints$ such that $\xi_\ell = \iroot{x_i}{p}{j}(s_{[i-1]})$}
    }
    \ElseIf{$s_i \in (\xi_k,\infty)$}{
        \Return{$(\iroot{x_i}{p}{j},\infty)$ for $p \in P$, $j \in \posints$ such that $\xi_k = \iroot{x_i}{p}{j}(s_{[i-1]})$}
    }
\end{algorithm}

\begin{algorithm}[b]
    \caption{\protect\FComputeCellProjection{$P$,$s$,$\Isymb$}.}
    \label{alg:cell:compute-cell-projection}

    \Input{$P \subseteq \rationals[x_1,\ldots,x_i]$, $s \in \reals^i$, symbolic interval $\Isymb$ of level $i$ for an $i \in \prange{n}$.}
    \Output{$P' \subseteq \rationals[x_1,\ldots,x_{i{-}1}]$ such that $\cell{P',s_{[i-1]}} \times \Isymb \subseteq \cell{P,s}$.}
\end{algorithm}

\noindent \Cref{alg:cell:construct} computes such a single cell using the levelwise algorithm from \cite{nalbach2024levelwise}. It uses \Cref{alg:cell:choose-interval} to determine the symbolic interval $\Isymb$ on a given level: by isolating the real roots of the polynomials, it determines which polynomial bounds the cell on that level. It then uses \Cref{alg:cell:compute-cell-projection} to compute projection polynomials that characterize an underlying cell where the bounds described by $\Isymb$ remain valid.
For $P \subseteq \rationals[x_1,\ldots,x_i]$ and $s \in \reals^i$, let $\cell{P,s} \subseteq \reals^i$ denote the inclusion-maximal $P$-sign-invariant cell that contains $s$.
Then due to its local cylindricity, the resulting cell $\Isymb_1 \times \ldots \times \Isymb_n$ will be a subset of $\cell{P,s}$. For the details of \Cref{alg:cell:compute-cell-projection}, we refer to \cite[Algorithm 2]{nalbach2024levelwise} and \cite[Algorithm 7]{nalbach2024calc}.

\begin{example}\label{ex:scc}
    \begin{figure}[t]
        \begin{subfigure}[t]{0.24\textwidth}
            \includegraphics[width=\textwidth]{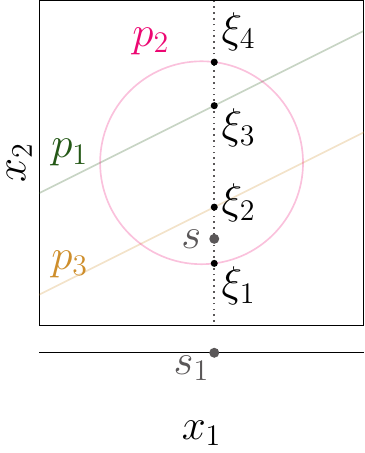}
            \caption{The sample point depicted on the one- and two-dimensional coordinate systems.
            }
            \label{fig:cells:levelwise:01}
        \end{subfigure}\hfill
        \begin{subfigure}[t]{0.24\textwidth}
            \includegraphics[width=\textwidth]{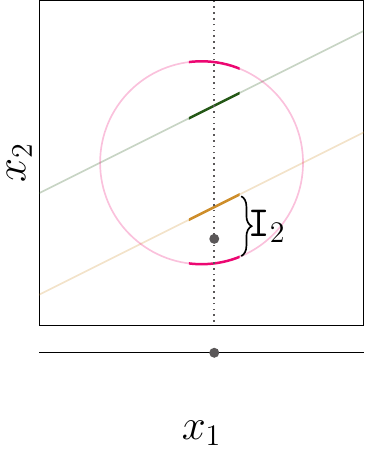}
            \caption{The real roots witness root functions of the polynomials.
            }
            \label{fig:cells:levelwise:02}
        \end{subfigure}\hfill
        \begin{subfigure}[t]{0.24\textwidth}
            \includegraphics[width=\textwidth]{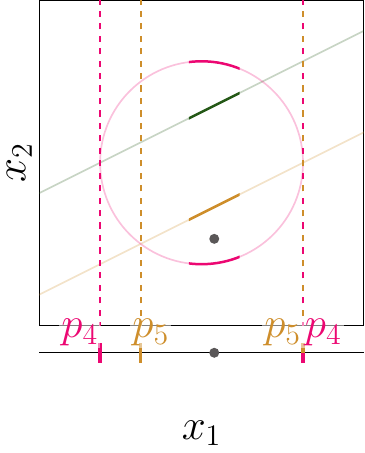}
            \caption{Zeros of projection polynomials define the symbolic interval bound.%
             }
            \label{fig:cells:levelwise:03}
        \end{subfigure}\hfill
        \begin{subfigure}[t]{0.24\textwidth}
            \includegraphics[width=\textwidth]{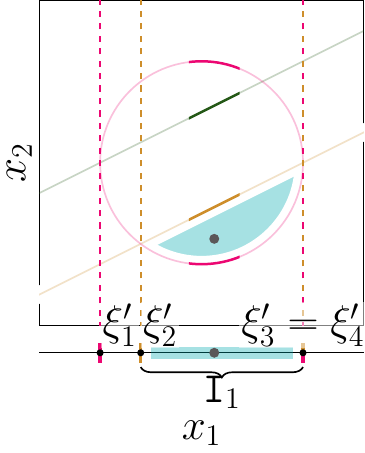}
            \caption{On the level below, we iterate the process.
            }
            \label{fig:cells:levelwise:04}
        \end{subfigure}
        \caption{Levelwise construction of the single cell from \Cref{ex:scc}.}
        \label{fig:cells:levelwise}
    \end{figure}

    Consider the polynomials $p_1 = 0.5 x_1 + 0.5-x_2$, $p_2 = x_1^2+x_2^2-1$, $p_3 = 0.5 x_1 - 0.5 -x_2$, and the point $s = (0.125,-0.75)$. 

    We consider the first coordinate $s_{1}=0.125$ of $s$, evaluate the polynomials at this value for $x_1$, and isolate the real roots of the resulting univariate polynomials to find $\xi_1=\iroot{x_2}{p_2}{1}$, $\xi_2=\iroot{x_2}{p_3}{1}$, $\xi_3=\iroot{x_2}{p_1}{1}$, and $\xi_4=\iroot{x_2}{p_2}{2}$.  We order these, along with the second coordinate of $s$, as in \Cref{fig:cells:levelwise:01}.  
   The symbolic interval containing $s$ is denoted $\Isymb_2=(\xi_1, \xi_2)$ in \Cref{fig:cells:levelwise:02}. Thus local to $s_{1}$, the cell we want is bounded from below by $p_2$ and from above by $p_3$.

    As we generalize from $s$ we must ensure that the domains of $\xi_1$ and $\xi_2$ remain well-defined, no root function crosses the symbolic interval $\Isymb_2$, and no additional roots pop up within $\Isymb_2$. To achieve this, we compute the projection polynomials $p_4$ and $p_5$ (details omitted here). \Cref{fig:cells:levelwise:03} shows how the zeros of the projection map onto the geometric features.
    Analogously to $x_2$, we isolate the real roots $\xi'_1 = \iroot{x_1}{p_4}{1}, \xi'_2 = \iroot{x_1}{p_5}{1}, \xi'_3 = \iroot{x_1}{p_4}{2}, \xi'_4 = \iroot{x_1}{p_5}{2}$ in $x_1$.  We determine the interval $\Isymb_1 = (\xi'_2, \xi'_3)$ around $s_{1}$ for $x_1$ and thus generate the cell in \Cref{fig:cells:levelwise:04}.\hfill\eoe
\end{example}

\section{Non-uniform Cylindrical Algebraic Decomposition}
\label{sec:nucad}

The NuCAD algorithm builds --- similarly to NLSAT and CAlC --- on top of the single cell construction. It computes a \emph{decomposition} of $\reals^n$, whose cells, however, are not \emph{globally} cylindrically arranged. It can be seen as a cylindrical decomposition, where some cells are further refined into a \emph{local} cylindrical decomposition.

\begin{definition}[Non-uniform CAD \cite{brown2015nucad}]
    Let $i \in \prange{n}$, and $R \subseteq \reals^i$ an algebraic cell.
    A \emph{non-uniform cylindrical algebraic decomposition (NuCAD) of $R$} is an algebraic decomposition $D$ of $R$ that consists of locally cylindrical cells such that there exists a partition $P_1,\ldots,P_k$ of $D$ (that is $P_1 \cup \ldots \cup P_k = D$ is a disjoint union) such that $\{ \cup_{C \in P_j} C \mid j \in \prange{k} \}$ is cylindrical, and for each $j \in \prange{k}$ the set $P_j$ consists either of a single element or is a NuCAD of $\cup_{C \in P_j} C$.\hfill\eod
\end{definition}

\begin{figure}[t]
    \centering
    \includegraphics[width=0.49\textwidth]{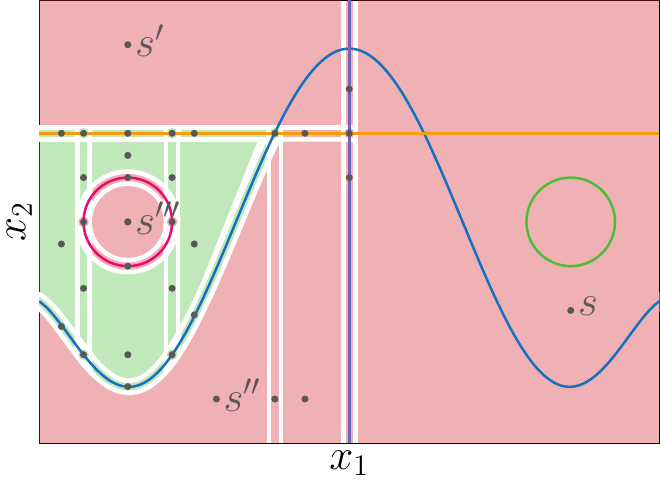}
\vspace*{-1ex}
    \caption{Truth-invariant non-uniform CAD, with a sample point for each cell.}
    \label{fig:nucad}
\end{figure}

\begin{example}\label{ex:nucad}
    We consider the formula from \Cref{ex:cad} again. A truth-invariant NuCAD for the input formula is depicted in \Cref{fig:nucad}.\hfill\eoe
\end{example}

\noindent We can represent a NuCAD by the following data structure:

\begin{definition}[NuCAD Data Structure]
    A \emph{NuCAD structure of level $i \in \prange{n}$ w.r.t. global level $i' \in \range{i}{n}$} is a finite sequence of pairs $(\Isymb_1, \Tsymb_1),\ldots,(\Isymb_k, \Tsymb_k)$ with $k \in \posints$ and for all $j \in \prange{k}$, $\Isymb_j$ is a symbolic interval of level $i$ and $\Tsymb_j$ is either $\true$, $\false$, a NuCAD structure of level $i+1 \leq i'$ w.r.t. global level $i'$, or a NuCAD structure of level $1$ w.r.t. global level $i'$.\hfill\eod
\end{definition}

\noindent We omit a formal semantics due to the space limit. Intuitively, each subtree of some node either describes a decomposition of the cylinder above the cell described by the intervals on the path to the given node (as long as the levels are increasing), or describes a decomposition of that cell (if the level is reset to $1$). We represent the NuCAD recursively to retain information about the arrangement of the cells, similar to \cite[Definition 2]{brown2015nucad}. The technicalities of this definition are not crucial for the NuCAD algorithm. We thus refrain from going into more detail.

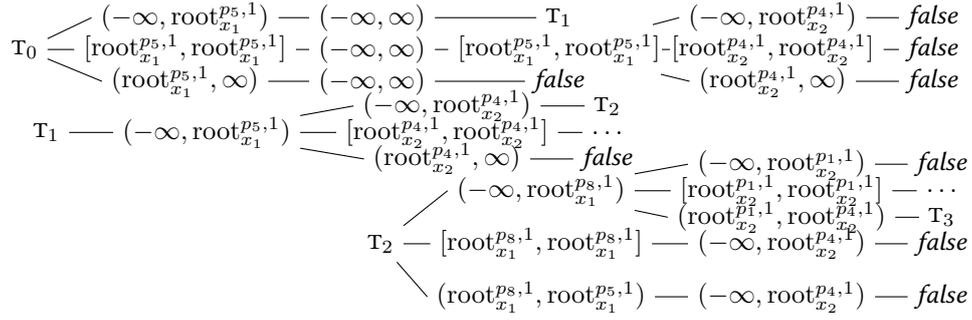
\begin{figure}[t]
    \centering
    \begin{tikzpicture}
        \node at (0,0) {
        \begin{tikzpicture}
            [grow'=right, level/.style={sibling distance=1.2em}, level 1/.style={level distance = 6em}, level 2/.style={level distance = 7em}, level 3/.style={level distance = 7em}, level 4/.style={level distance = 8em}, level 5/.style={level distance = 6em}, child anchor=west]
            \node{$\Tsymb_0$}
            child {node {$(-\infty,\iroot{x_1}{p_5}{1})$}
                child {node {$(-\infty,\infty)$}
                child {node {$\Tsymb_1$}}
                }
            }
            child {node {$[\iroot{x_1}{p_5}{1},\iroot{x_1}{p_5}{1}]$}
                child {node {$(-\infty,\infty)$}
                child {node {$[\iroot{x_1}{p_5}{1},\iroot{x_1}{p_5}{1}]$}
                    child {node {$(-\infty,\iroot{x_2}{p_4}{1})$}
                    child {node {$\false$}}
                    }
                    child {node {$[\iroot{x_2}{p_4}{1},\iroot{x_2}{p_4}{1}]$}
                    child {node {$\false$}}
                    }
                    child {node {$(\iroot{x_2}{p_4}{1},\infty)$}
                    child {node {$\false$}}
                    }
                }
                }
            }
            child {node {$(\iroot{x_1}{p_5}{1},\infty)$}
                child {node {$(-\infty,\infty)$}
                child {node {$\false$}}
                }
            }
            ;
        \end{tikzpicture}};
        \node at (-2,-1.1) {
        \begin{tikzpicture}
            [grow'=right, level/.style={sibling distance=10em/(#1*#1*#1+2)}, level 1/.style={level distance = 6em}, level 2/.style={level distance = 9em}, level 3/.style={level distance = 6em}, child anchor=west]
            \node{$\Tsymb_1$}
            child {node {$(-\infty,\iroot{x_1}{p_5}{1})$}
                child {node {$(-\infty,\iroot{x_2}{p_4}{1})$}
                child {node {$\Tsymb_2$}}
                }
                child {node {$[\iroot{x_2}{p_4}{1},\iroot{x_2}{p_4}{1}]$}
                child {node {$\ldots$}}
                }
                child {node {$(\iroot{x_2}{p_4}{1},\infty)$}
                child {node {$\false$}}
                }
            }
            ;
        \end{tikzpicture}};
        \node at (2.4,-2.4) {
        \begin{tikzpicture}
            [grow'=right, level/.style={sibling distance=10em/(#1*#1*#1+2)}, level 1/.style={level distance = 6em, sibling distance=2em}, level 2/.style={level distance = 9em}, level 3/.style={level distance = 6em}, child anchor=west]
            \node{$\Tsymb_2$}
            child {node {$(-\infty,\iroot{x_1}{p_8}{1})$}
                child {node {$(-\infty,\iroot{x_2}{p_1}{1})$}
                child {node {$\false$}}
                }
                child {node {$[\iroot{x_2}{p_1}{1},\iroot{x_2}{p_1}{1}]$}
                child {node {$\ldots$}}
                }
                child {node {$(\iroot{x_2}{p_1}{1},\iroot{x_2}{p_4}{1})$}
                child {node {$\Tsymb_3$}}
                }
            }
            child {node {$[\iroot{x_1}{p_8}{1},\iroot{x_1}{p_8}{1}]$}
                child {node {$(-\infty,\iroot{x_2}{p_4}{1})$}
                child {node {$\false$}}
                }
            }
            child {node {$(\iroot{x_1}{p_8}{1},\iroot{x_1}{p_5}{1})$}
                child {node {$(-\infty,\iroot{x_2}{p_4}{1})$}
                child {node {$\false$}}
                }
            }
            ;
        \end{tikzpicture}};
            \end{tikzpicture}
    \caption{NuCAD data structure that represents the NuCAD in \Cref{fig:nucad}.}
    \label{fig:nucadds}
\end{figure}

\begin{figure}[t]
    \begin{subfigure}{0.49\textwidth}
        \includegraphics[width=\textwidth]{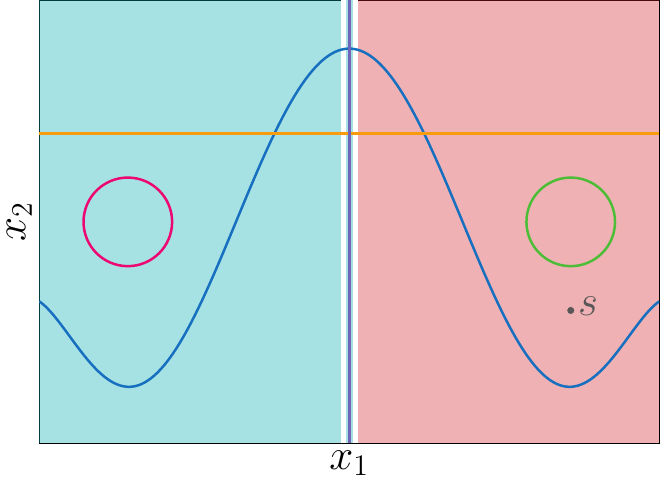}

        \vspace*{-1ex}
        \caption{Partition represented by $\Tsymb_0$.}
        \label{fig:nucad:steps:01}
    \end{subfigure}\hfill
    \begin{subfigure}{0.49\textwidth}
        \includegraphics[width=\textwidth]{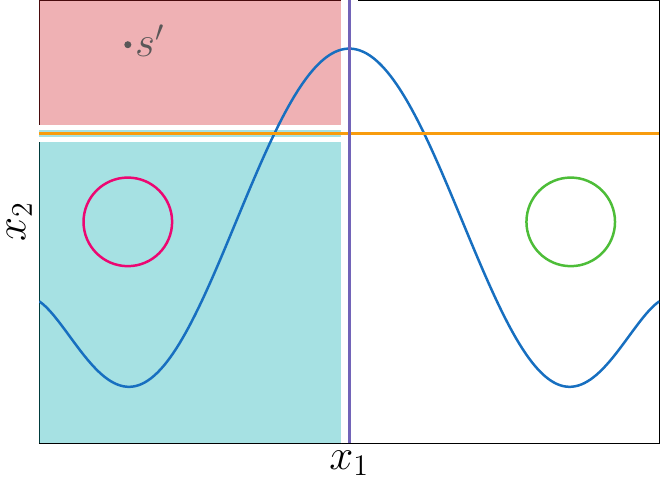}

        \vspace*{-1ex}
        \caption{Partition represented by $\Tsymb_1$.}
        \label{fig:nucad:steps:02}
    \end{subfigure}
    \begin{subfigure}{0.49\textwidth}
        \includegraphics[width=\textwidth]{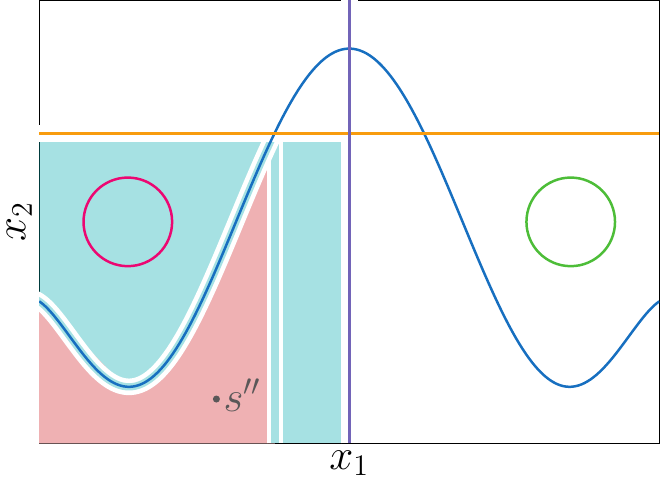}

        \vspace*{-1ex}
        \caption{Partition represented by $\Tsymb_2$.}
        \label{fig:nucad:steps:03}
    \end{subfigure}\hfill
    \begin{subfigure}{0.49\textwidth}
        \includegraphics[width=\textwidth]{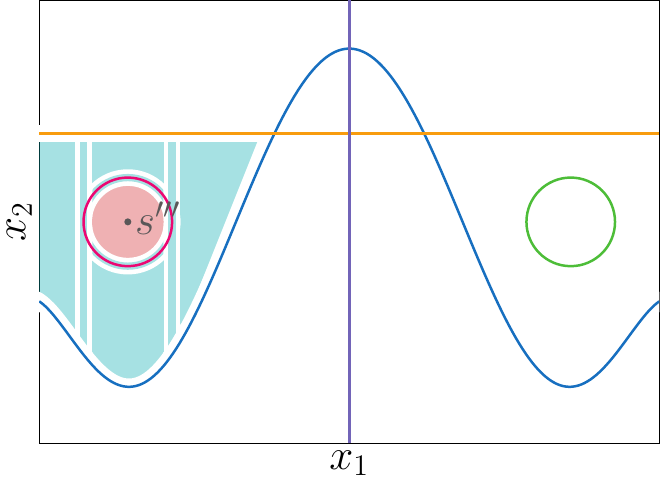}

        \vspace*{-1ex}
        \caption{Partition represented by $\Tsymb_3$.}
        \label{fig:nucad:steps:04}
    \end{subfigure}
    \caption{Some partitions of cells in the NuCAD from \Cref{fig:nucad,fig:nucadds}.}
    \label{fig:nucad:steps}
\end{figure}

\begin{example}
    The NuCAD depicted in \Cref{fig:nucad} is represented by the tree partially depicted in \Cref{fig:nucadds}. A path from the root of length $2$ describes a cell of the decomposition, and the following subtree the decomposition of that cell, e.g., $\Tsymb_0$ is the tree decomposing $\reals^2$, $\Tsymb_1$ is the decomposition of the first cell $(-\infty,\iroot{x_1}{p_5}{1}) \times (-\infty,\infty)$ of $\Tsymb_1$. \Cref{fig:nucad:steps} displays the corresponding decompositions.
    \hfill\eoe
\end{example}

\subsection{A Complete Algorithm for Computing a NuCAD}
\label{sec:nucad:algo}

We present a version of the algorithm from \cite[Algorithm 2]{brown2015} for NuCAD computation, which we adapted to be \emph{complete} (i.e. not restricted to open cells only).
To handle the Boolean structure of the formula, we make use of \Cref{alg:calc:choose-enclosing-cell} for computing implicants as in \Cref{def:implicant}. For details, we refer to \cite[Section 6]{nalbach2024calc}.

\SetKwFunction{FChooseEnclosingCell}{choose\_enclosing\_cell}
\begin{algorithm}[b]
    \caption{\protect\FChooseEnclosingCell{$\varphi$, $s$}.}
    \label{alg:calc:choose-enclosing-cell}
	\Input{Formula $\varphi$, $s \in \reals^{i}$ for an $i \in \prange{n}$ s.t. $\varphi[s] \equiv \false$ or $\varphi[s] \equiv \true$.}
	\Output{$P \subseteq \rationals[x_1,\ldots,x_i]$, $\texttt{t} \in \{ \true,\false \}$ s.t. $\varphi(s') \equiv \texttt{t}$ for all $s' \in C(P,s)$.}  
\end{algorithm}

\SetKwFunction{FComputeNucadRecurse}{nucad\_recurse}
\SetKwFunction{FComputeNucadFull}{nucad\_full}
\SetKwFunction{FComputeNucadQuantifier}{nucad\_quantifier}

\begin{algorithm}[t]

  \caption{\protect\FComputeNucadFull{$\varphi$, $\hlmath{s,}$ $\hlremove{(\Isymb_{1},\ldots,\Isymb_{n})}$ $\hlmath{(\Isymb_{i{+}1},\ldots,\Isymb_{i'})}$}.}
    \label{alg:nucad:compute-nucad-full}

    \Input{Formula $\varphi$, $\hlmath{s \in \reals^i \text{ for an } i \in \zrange{n}}$, \\
      symbolic intervals $\hlremove{(\Isymb_{1},\ldots,\Isymb_{n})}$
      $\hlmath{(\Isymb_{i{+}1},\ldots,\Isymb_{i'}) \text{ for an } i' \in \range{i{+}1}{n}}$.}
    \Output{$\hlmath{P \subseteq \rationals[x_1,\ldots,x_i]}$, and NuCAD data structure $\Tsymb$ representing a truth-invariant NuCAD of\\
      $\hlremove{\Isymb_{1}\times\ldots\times\Isymb_{n}}$
      $\hlmath{\cell{P,s} \times \Isymb_{i{+}1}\times\ldots\times\Isymb_{i'} \textit{ over } \cell{P,s}}$.}

    \textbf{choose}
    \hlremove{$r \in \reals^{n}$ s.t. $r \in \Isymb_{1} \times \ldots \times \Isymb_{n}$}
    \hlalgo{$r \in \reals^{i'}$ s.t. $r \in s \times \Isymb_{i{+}1} \times \ldots \times \Isymb_{i'}$}
    \; \label{alg:nucad:compute-nucad-full:sample}

    $P$, $\texttt{t}$ $ := $ \hlremove{\FChooseEnclosingCell{$\varphi$, $r$}} \hlalgo{\FComputeNucadRecurse{$\varphi$, $r$}} \; \label{alg:nucad:compute-nucad-full:recurse} %

    $P := P \cup \{ \Isymb_j.l.p \mid j \in
    \hlremove{\range{1}{n}} \ \hlmath{\range{i{+}1}{i'}},
    \Isymb_j.l {\neq} -\infty \} \cup \newline
    \phantom{:=P \cup \{} \{ \Isymb_j.u.p \mid j \in
    \hlremove{\range{1}{n}} \ \hlmath{\range{i{+}1}{i'}},
    \Isymb_j.u {\neq} \infty \}$\; \label{alg:nucad:compute-nucad-full:input}
    \ForEach{$j = \hlremove{n,\ldots,1} \ \hlmath{i',\ldots,i{+}1}$}{ \label{alg:nucad:compute-nucad-full:scc}
        $\Isymb'_j := $ \FChooseInterval{$P$,$r_{[j]}$} \tcp*{\Cref{alg:cell:choose-interval}}
        $P := $ \FComputeCellProjection{$P$,$r_{[j]}$,$\Isymb'_j$} \tcp*{\Cref{alg:cell:compute-cell-projection}}
    }

    $\Tsymb := \emptyset$ \;
    insert $\hlremove{\Isymb'_{1},\ldots,\Isymb'_{n}} \ \hlmath{\Isymb'_{i{+}1},\ldots,\Isymb'_{i'}} $ into $\Tsymb$ leading to  $\texttt{t}$ \; \label{alg:nucad:compute-nucad-full:insert}

    $Q := \emptyset$ \;
    \ForEach{$j=\hlremove{1,\ldots,n} \ \hlmath{i{+}1,\ldots,i'}$}{ \label{alg:nucad:compute-nucad-full:splitting}
        \uIf{$\Isymb'_j.l \neq \Isymb_j.l$}{
            $Q := Q \cup \{ (\hlremove{\Isymb'_{1}} \ \hlmath{\Isymb'_{i{+}1}} , \ldots, \Isymb'_{j-1}, (\Isymb_j.l,\Isymb'_j.l), \Isymb_{j+1}, \ldots, \hlremove{\Isymb_{n}}\ \hlmath{\Isymb_{i'}}) \}$ \;
            \uIf{$\Isymb'_j$ is a sector}{
                $Q := Q \cup \{ (\hlremove{\Isymb'_{1}} \ \hlmath{\Isymb'_{i{+}1}}, \ldots, \Isymb'_{j-1}, [\Isymb'_j.l,\Isymb'_j.l], \Isymb_{j+1}, \ldots, \hlremove{\Isymb_{n}}\ \hlmath{\Isymb_{i'}}) \}$
            }
        }
        \uIf{$\Isymb'_j.u \neq \Isymb_j.u$}{
            $Q := Q \cup \{ (\hlremove{\Isymb'_{1}} \ \hlmath{\Isymb'_{i{+}1}}, \ldots, \Isymb'_{j-1}, (\Isymb'_j.u,\Isymb_j.u), \Isymb_{j+1}, \ldots, \hlremove{\Isymb_{n}}\ \hlmath{\Isymb_{i'}}) \}$ \;
            \uIf{$\Isymb'_j$ is a sector}{
                $Q := Q \cup \{ (\hlremove{\Isymb'_{1}} \ \hlmath{\Isymb'_{i{+}1}}, \ldots, \Isymb'_{j-1}, [\Isymb'_j.u,\Isymb'_j.u], \Isymb_{j+1}, \ldots, \hlremove{\Isymb_{n}}\ \hlmath{\Isymb_{i'}}) \}$
            }
        }
        \label{alg:nucad:compute-nucad-full:splittingend}
    }

    \ForEach{$\hlremove{(\Isymb'_{1},\ldots,\Isymb'_{n})} \ \hlmath{(\Isymb'_{i{+}1},\ldots,\Isymb'_{i'})} \in Q$}{
        $\hlmath{P'}, \Tsymb' := $ \FComputeNucadFull{$\varphi$, $\hlmath{s}$, $\hlremove{(\Isymb'_{1},\ldots,\Isymb'_{n})} \ \hlmath{(\Isymb'_{i{+}1},\ldots,\Isymb'_{i'})}$} \; \label{alg:nucad:compute-nucad-full:subnucad}
        $\hlmath{P := P \cup P'}$ \; \label{alg:nucad:compute-nucad-full:projection}
        insert $\hlremove{\Isymb'_{1},\ldots,\Isymb'_{n}} \ \hlmath{\Isymb'_{i{+}1},\ldots,\Isymb'_{i'}}$ into $\Tsymb$ leading to $\Tsymb'$ \; \label{alg:nucad:compute-nucad-full:insert2}
    }

    \Return{$\hlmath{P}$, $\Tsymb$}
\end{algorithm}

Let us assume that we generate a NuCAD of some cell $R = \Isymb_{1}\times\ldots\times\Isymb_{n} \subseteq \reals^n$ for a quantifier-free formula $\varphi$ using \Cref{alg:nucad:compute-nucad-full}; for now, ignore the highlighted (but consider the underlined) bits in the algorithm. We start by picking a sample point contained in $R$ in \Cref{alg:nucad:compute-nucad-full:sample}, and in \Cref{alg:calc:choose-enclosing-cell}, we obtain an implicant for $\varphi$ w.r.t. $s$. In \Cref{alg:nucad:compute-nucad-full:scc}, we apply the single cell construction algorithm on the obtained implicit cell. To compute a cell that is a subset of $R \subseteq \reals^n$, we insert the defining polynomials of $\Isymb_{1},\ldots,\Isymb_{n}$ to $P$ in \Cref{alg:nucad:compute-nucad-full:input} before constructing the cell.

The obtained cell $R' = \Isymb'_{1}\times\ldots\times\Isymb'_{n}$ is thus a subset of $R$, and $\varphi$ is truth-invariant on $R'$. We insert the cell into the NuCAD data structure $\Tsymb$ in \Cref{alg:nucad:compute-nucad-full:insert}, such that --- when seen as a tree --- there is a path $\Isymb'_{1},\ldots,\Isymb'_{n}$ from the root node leading to $\texttt{t}$. What remains is to explore the remaining parts $R \setminus R'$. We do so by splitting $R$ using $R'$ into multiple cylindrical cells in \Cref{alg:nucad:compute-nucad-full:splitting}: we observe that a point is outside the cell if it violates one bound. We thus iterate through each level $i$ and encode that the first $i-1$ levels are in $R'$; on the $i$-th level, we leave $R'$ (we are either below or above the cell, if possible) but stay in $R$; and on the levels starting from $i+1$, we remain in $R$, but do not make assumptions about the bounds on $R'$, as we already left $R'$ on the $i$-th level.

We end up with a set $Q$ of unexplored cells which are locally cylindrical, i.e. they are described by symbolic intervals for each variable. We call the NuCAD algorithm on each cell in $Q$ in \Cref{alg:nucad:compute-nucad-full:subnucad} to explore the remaining parts of $R$.%

\begin{example}
    Consider the formula $\varphi(x_1,x_2) = p_1 \leq 0 \wedge p_2 > 0 \wedge p_3 \geq 0 \wedge p_4 \leq 0 \wedge p_5 \leq 0$ and the polynomials from \Cref{ex:cad} again. We give an exemplary run of \Cref{alg:nucad:compute-nucad-full} that computes the NuCAD as depicted in \Cref{fig:nucad,fig:nucadds}.

    \begin{description}
        \item[\FComputeNucadFull{$\varphi$, $((-\infty,\infty),(-\infty,\infty))$}]
        We choose $s$ as sample point. As $\varphi$ evaluates to $\false$ at $s$, we obtain the implicant $p_5>0$, and thus obtain $(\{p_5\},s)$ as implicit cell. By single cell construction, we obtain the cell described by $\Isymb'_1\times\Isymb'_2=(\iroot{x_1}{p_5}{1},\infty)\times(-\infty,\infty)$. Accordingly, we split $(-\infty,\infty)\times(-\infty,\infty)$ into $(-\infty,\iroot{x_1}{p_5}{1})\times(-\infty,\infty)$, $[\iroot{x_1}{p_5}{1},\iroot{x_1}{p_5}{1}]\times(-\infty,\infty)$, and $\Isymb'_1\times\Isymb'_2$, as depicted in \Cref{fig:nucad:steps:01}. 
        
        \begin{description}
            \item[\FComputeNucadFull{$\varphi$, $((-\infty,\iroot{x_1}{p_5}{1}),(-\infty,\infty))$}] 
            We choose the sample point $s'$, and obtain the implicit cell $(\{p_4\},s')$ where $\varphi$ is truth-invariant. As we aim to compute a truth-invariant cell that is a subset of $(-\infty,\iroot{x_1}{p_5}{1})\times(-\infty,\infty)$, we run single cell construction on $(\{ p_4,p_5 \}, s')$. We obtain $\Isymb'_1\times\Isymb'_2=(-\infty,\iroot{x_1}{p_5}{1})\times(\iroot{x_2}{p_4}{1},\infty)$, and split into $(-\infty,\iroot{x_1}{p_5}{1})\times(-\infty,\iroot{x_2}{p_4}{1})$,$(-\infty,\iroot{x_1}{p_5}{1})\times[\iroot{x_2}{p_4}{1},\iroot{x_2}{p_4}{1}]$, and $\Isymb'_1\times\Isymb'_2$, as depicted in \Cref{fig:nucad:steps:02}.

            \begin{description}
                \item[\FComputeNucadFull{$\varphi$, $((-\infty,\iroot{x_1}{p_5}{1}),(-\infty,\iroot{x_2}{p_4}{1}))$}] 
                We choose $s''$, and obtain $(\{p_1\},s'')$ from the implicant. We run single cell construction on $(\{p_1,p_4,p_5\},s'')$ and obtain $(-\infty,\iroot{x_1}{p_8}{1})\times(-\infty,\iroot{x_2}{p_1}{1})$ (where $p_8$ is the resultant of $p_1$ and $p_4$), and split the input cell into the cells as indicated in \Cref{fig:nucad:steps:03} and the following recursive calls:

                \begin{description}
                    \item[\FComputeNucadFull{$\ldots$, $([\iroot{x_1}{p_8}{1},\iroot{x_1}{p_8}{1}],(-\infty,\iroot{x_2}{p_4}{1}))$}] $\ldots$
                    \item[\FComputeNucadFull{$\ldots$, $((\iroot{x_1}{p_8}{1},\iroot{x_1}{p_5}{1}),(-\infty,\iroot{x_2}{p_4}{1}))$}] $\ldots$
                    \item[\FComputeNucadFull{$\ldots$, $((-\infty,\iroot{x_1}{p_8}{1}),[\iroot{x_2}{p_1}{1},\iroot{x_2}{p_1}{1}])$}] $\ldots$
                    \item[\FComputeNucadFull{$\ldots$, $((-\infty,\iroot{x_1}{p_8}{1}),(\iroot{x_2}{p_1}{1},\iroot{x_2}{p_4}{1}))$}] 
                    
                    We choose $s''$ and obtain $(\{p_2\},s''')$ from the implicant. We run single cell construction on $(\{p_1,p_2,p_4\},s''')$ and obtain $(\iroot{x_1}{p_6}{1},\iroot{x_1}{p_6}{2})\times(\iroot{x_2}{p_2}{1},\iroot{x_2}{p_2}{2})$ (where $p_6$ is the resultant of $p_2$), and split the input cell into the cells as indicated in \Cref{fig:nucad:steps:04} and computed by some recursive calls (omitted due to limited space). %
                                    \end{description}

                \item[\FComputeNucadFull{$\varphi$, $((-\infty,\iroot{x_1}{p_5}{1}),[\iroot{x_2}{p_4}{1},\iroot{x_2}{p_4}{1}])$}] $\ldots$
            \end{description}

            \item[\FComputeNucadFull{$\varphi$, $([\iroot{x_1}{p_5}{1},\iroot{x_1}{p_5}{1}],(-\infty,\infty))$}] $\ldots$
        \end{description}
    \end{description}
    \vspace{-2em}
    \hfill\eoe
\end{example}

\begin{algorithm}[t]
    \caption{\protect\FComputeNucadRecurse{$\varphi$, $s$}.}
    \label{alg:nucad:compute-nucad-recurse}

	\Input{Formula $\varphi$ in prenex normal form, $s \in \reals^i$ for an $i \in \zrange{n}$ s.t. $\varphi$ contains at most $i$ free variables.}
	\Output{$P \subseteq \rationals[x_1,\ldots,x_i]$, %
          $\texttt{t} \in \{ \true,\false \}$ s.t. $\varphi(s') \equiv \texttt{t}$ for all $s' \in C(P,s)$.%
          }

	\uIf{$\varphi[s] \equiv \false \vee \varphi[s] \equiv \true$}{
		\Return{$\FChooseEnclosingCell{$\varphi$, $s$}$} \tcp*{\Cref{alg:calc:choose-enclosing-cell}}
	}
	\Else(it holds $i<n$){
		\uIf{$\varphi = \exists x_{i+1}\ldots\exists x_{i'}.\; \psi$}{
			\Return{\FComputeNucadQuantifier{$\exists$, $\psi$, $s$, $i'$, $(-\infty,\infty),\ldots,(-\infty,\infty)$}} \tcp*[f]{A \ref{alg:nucad:compute-nucad-quantifier}}
		}
		\uElseIf{$\varphi = \forall x_{i+1}\ldots\forall x_{i'}.\; \psi$}{
			\Return{\FComputeNucadQuantifier{$\forall$, $\psi$, $s$, $i'$, $(-\infty,\infty),\ldots,(-\infty,\infty)$}} \tcp*[f]{A \ref{alg:nucad:compute-nucad-quantifier}}
		}	
	}
\end{algorithm}

\section{Quantifier Alternation in NuCAD}
\label{sec:nucad:quantifiers}

\begin{algorithm}[tb]
    \caption{\protect\FComputeNucadQuantifier{$Q$, $\varphi$, $s$, $(\Isymb_{i{+}1},\ldots,\Isymb_{i'})$}.}
    \label{alg:nucad:compute-nucad-quantifier}

    \Input{$Q \in \{ \exists, \forall\}$, formula $\varphi$, $s \in \reals^i$ for an $i \in \zrange{n}$, \\ symbolic intervals $(\Isymb_{i{+}1},\ldots,\Isymb_{i'})$ for an $i' \in \range{i{+}1}{n}$.}
    \Output{$P \subseteq \rationals[x_1,\ldots,x_i]$, $\texttt{t} \in \{ \true,\false \}$ s.t. $\varphi(s') \equiv \texttt{t}$ for all $s' \in C(P,s)$.}

    \textbf{choose} $r \in \reals^{i'} $ s.t. $r \in s \times \Isymb_{i{+}1} \times \ldots \times \Isymb_{i'}$ \;

    $P$, $\texttt{t} := $ \FComputeNucadRecurse{$\varphi$, $r$} \tcp*{\Cref{alg:nucad:compute-nucad-recurse}}

    \uIf{$(Q = \exists \wedge \texttt{t} = \true) \vee (Q = \forall \wedge \texttt{t} = \false$)}{
        \ForEach{$j = {i'},\ldots,i{+}1$}{
            $\Isymb'_j := $ \FChooseInterval{$P$,$r_{[j]}$} \tcp*{\Cref{alg:cell:choose-interval}}
            $P := $ \FComputeCellProjection{$P$,$r_{[j]}$,$\Isymb'_j$} \tcp*{\Cref{alg:cell:compute-cell-projection}}
        }
        \Return{$P$, $\texttt{t}$} \label{alg:nucad:compute-nucad-quantifier:early}
    }

    $P := P \cup \{ \Isymb_j.l.p \mid j \in \range{i{+}1}{i'}, \Isymb_j.l \neq -\infty \} \cup \{ \Isymb_j.u.p \mid j \in \range{i{+}1}{i'}, \Isymb_j.u \neq \infty \}$ \;

    \ForEach{$j = {i'},\ldots,i{+}1$}{
        $\Isymb'_j := $ \FChooseInterval{$P$,$r_{[j]}$} \tcp*{\Cref{alg:cell:choose-interval}}
        $P := $ \FComputeCellProjection{$P$,$r_{[j]}$,$\Isymb'_j$} \tcp*{\Cref{alg:cell:compute-cell-projection}}
    }

    $Q := \emptyset$ \;
        \textbf{for}-loop in Lines~\ref{alg:nucad:compute-nucad-full:splitting}-\ref{alg:nucad:compute-nucad-full:splittingend} of \Cref{alg:nucad:compute-nucad-full}

    \ForEach{$(\Isymb'_{i{+}1},\ldots,\Isymb'_{i'}) \in Q$}{
        $P', \texttt{t} := $ \FComputeNucadQuantifier{$Q$, $\varphi$, $s$, $(\Isymb'_{i{+}1},\ldots,\Isymb'_{i'})$} \;

        \uIf{$(Q = \exists \wedge \texttt{t} = \true) \vee (Q = \forall \wedge \texttt{t} = \false)$}{
            \Return{$P'$, $\texttt{t}$} \label{alg:nucad:compute-nucad-quantifier:recearly}
        }

        $P := P \cup P'$ \;
    }

    \lIf{$Q = \exists$}{$\texttt{t} := \false$ \textbf{else} $\texttt{t} := \true$}
    \Return{$P$, $\texttt{t}$}
\end{algorithm}

A NuCAD as computed by the algorithm presented in the previous section is not eligible for reasoning about quantifier alternations, as it does not have the cylindrical structure. An extension for quantifier elimination is given in \cite{brown2017nucad}, where a NuCAD of $\reals^n$ is refined such that reasoning about quantifiers is possible.

Borrowing ideas from CAlC, we present an adaptation which directly computes a NuCAD adapted to the input's quantifier structure, making the refinement step obsolete. For doing so, we define a NuCAD over a cell, which allows to generalize the notion of cylindricity from single levels to blocks of levels.

\begin{definition}
    Let $i \in \zrange{n}$, $i' \in \range{i{+}1}{n}$, and $R \subseteq \reals^{i'}$ be an algebraic cell. A \emph{NuCAD of $R$ over $\proj{R}{\prange{i}}$} is a NuCAD $D$ of $R$ s.t. $\proj{R'}{\prange{i}} = \proj{R}{\prange{i}}$ for all $R' \in D$.\hfill\eod
\end{definition}

If we call \Cref{alg:nucad:compute-nucad-full} --- now we consider the highlighted and ignore the underlined parts --- with a sample point $s \in \reals^i$ and $i'<n$, it will compute a NuCAD over some cell around $s$ of $\reals^{i'}$. For doing so, it fixes the first $i$ levels of the guessed sample point in \Cref{alg:nucad:compute-nucad-full:sample}, applies single cell construction only from level $i'$ to level $i+1$ in \Cref{alg:nucad:compute-nucad-full:scc}, applies the splitting accordingly in \Cref{alg:nucad:compute-nucad-full:splitting}, and --- most importantly for the correctness --- collects the polynomials that characterize the NuCAD. This set consists of the polynomials of level $i$ and below that are left over from the single cell construction, and the projections obtained from the recursive calls to NuCAD (see \Cref{alg:nucad:compute-nucad-full:projection}). Further, as we do not compute a NuCAD of $\reals^n$ but of $\reals^{i'}$, the call to \Cref{alg:nucad:compute-nucad-recurse} in \Cref{alg:nucad:compute-nucad-full:recurse} gets important, as it obtains a truth-invariant cell for $\varphi$ even if $s$ is only a partial sample point.

\Cref{alg:nucad:compute-nucad-recurse} decides whether to call \Cref{alg:calc:choose-enclosing-cell} or the NuCAD algorithm (we could replace \FComputeNucadQuantifier{...} by \FComputeNucadFull{...}). For handling formulas with quantifiers, it ``structures'' the NuCAD such that we construct a NuCAD for each quantifier block separately.

Although we could use \Cref{alg:nucad:compute-nucad-full} for quantifier blocks, we only use it for describing the parameter space (by choosing $i'$ as the highest index of a free variable in the input formula); for quantifier blocks, similarly to the CAlC algorithm, we can terminate the NuCAD construction earlier, depending on the quantifier: if the given variables are existentially quantified, we can stop when we find a cell where the input formula is $\true$, and return a projection of that cell (instead of the NuCAD). Analogously, if the given variables are universally quantified, we stop when we find a cell where the input formula is $\false$. \Cref{alg:nucad:compute-nucad-quantifier} implements this idea and differs from \Cref{alg:nucad:compute-nucad-full} only in two places: if a call to \Cref{alg:nucad:compute-nucad-recurse} yields a desired cell, we call single cell construction and return the result in \Cref{alg:nucad:compute-nucad-quantifier:early}. If a recursive call to itself yields such a cell, we return that cell in \Cref{alg:nucad:compute-nucad-quantifier:recearly}, omitting the intermediate results from the NuCAD computation.

\section{Experimental Evaluation}
\label{sec:experiments}

We implemented the \texttt{NuCAD} algorithm in \texttt{SMT-RAT} \cite{corzilius2015}, which also provides a \texttt{CAlC} implementation.
Both implementations are sequential and share large portions of code (e.g. for Boolean reasoning as described in \cite{nalbach2024calc}, CAD projection as described in 
\cite{nalbach2024levelwise}, the variable ordering from \cite{pickering2024}, an adaption for the projection in \cite{nalbach2024mergingcells} described in \Cref{sec:nucad:improved}, and post-processing as described in \Cref{sec:nucad:encoding}), increasing their comparability. We run our experiments on Intel®Xeon®8468 Sapphire CPUs with 2.1 GHz per core with a time limit of 60 seconds and memory limit of 4GB per formula. For the evaluation, we use all instances from the \emph{QF\_NRA} and \emph{NRA} benchmark sets from \emph{SMT-LIB} \cite{preiner2024smtlib} for quantifier-free and quantified decision problems, and a collection of quantifier elimination \emph{(\!QE\!)} problems from Wilson \cite{wilson2013}. The code, raw results and evaluation scripts are available at \href{https://doi.org/10.5281/zenodo.15302831}{\url{https://doi.org/10.5281/zenodo.15302831}}.
We verified all results of our implementations using \texttt{QEPCAD B}/\texttt{Tarski} for quantified formulas, and using the MCSAT implementation of \texttt{SMT-RAT} for quantifier-free formulas, as they accept indexed root expressions as input.

\subsection{Comparison of NuCAD and CAlC}

\begin{figure}[tb]
    \centering
    \includegraphics[scale=0.7]{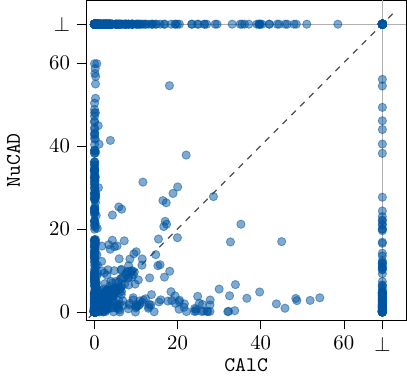}
    \caption{Running time in seconds. $\bot$ indicates a timeout.}
    \label{fig:results_qeqfnra}
\end{figure}

\begin{table}[tb]
    \caption{Statistics of the solvers on instances solved by both solvers. Each row is the sum of the given statistic of the given instances.}
    \label{fig:stats}
    \begin{tabular}{lp{15mm}p{15mm}}\toprule
        & \texttt{CAlC} & \texttt{NuCAD} \\ \midrule
        \# atoms & 93\,118 & 142\,668 \\
        \# cells & 123\,696 & 166\,819 \\
        \# symbolic intervals & 173\,378 & 727\,891 \\
        \# sections & 100\,492 & 3\,003\,788 \\
        real root time (s) & 70 ($17.59\%$) & 4652 ($61\%$) \\
        non-alg. time (s) & 3141 ($78.39\%$) &  2738 ($35.64\%$) \\
        \bottomrule
    \end{tabular}
\end{table}

As the NRA and QE benchmark sets do mostly contain small formulas, we compare \texttt{NuCAD} and \texttt{CAlC} on the QF\_NRA benchmark set by treating all variables as parameters, e.g. we generate a formula that describes all solutions by quantifier elimination. Note that the input formula already does this implicitly; however, the results by quantifier elimination describe them by cylindrical cells, which makes e.g. the generation of a solution point straight-forward. 
Of the $12154$ instances, \texttt{CAlC} solved $9559$, while \texttt{NuCAD} only solved $9253$; \texttt{CAlC} solved $422$ not solved by \texttt{NuCAD}, while \texttt{NuCAD} solved $116$ not solved by \texttt{CAlC}. \Cref{fig:results_qeqfnra} shows that \texttt{CAlC} is much faster on the majority of the instances, but there are instances where \texttt{NuCAD} is more efficient. Thus, the algorithms seem complementary to some degree.

Looking at \Cref{fig:stats}, \texttt{CAlC} produces smaller solution formulas. Comparing the number of cells (before the simplification) and symbolic intervals that describe the cells, we see that \texttt{NuCAD} requires more cells and proportionally more computation steps (reflected by the number of symbolic intervals).
The reason for this clear difference is not obvious: as the CAlC algorithm computes coverings of cylinders (which may lead to repeated expensive computations), it may visit the same cells multiple times, while NuCAD visits each cell only once (sometimes leading to coarser decompositions, e.g. compare \Cref{fig:calc} and \Cref{fig:nucad}).

Further, the \texttt{CAlC} avoids the costly exploration of sections more successfully than \texttt{NuCAD}: Looking again at \Cref{fig:stats}, \texttt{NuCAD} explores $29$ times more sections than \texttt{CAlC}. As consequence, \texttt{NuCAD} spends significantly more time on real root isolation.
We hoped that the optimization in \Cref{sec:nucad:improved} might avoid large parts of the sections. However, on instances solved by both solvers, during the run of \texttt{NuCAD}, only $5\%$ of all non-point intervals have at least one closed bound. For \texttt{CAlC}, which implements a similar technique, this number is $90\%$.

Still, \texttt{NuCAD} spends fewer time on non-algebraic computations. In fact, we observe that \texttt{NuCAD} is able to solve bigger formulas (the mean number of nodes in the directed acyclic graph representing the formula is $833$ ($1565$) over the instances solved exclusively by \texttt{NuCAD} (\texttt{CAlC})), while \texttt{CAlC} is able to solve formulas with higher degree (the mean maximal degree is $3.8$ ($9.85$) over the instances solved exclusively by \texttt{NuCAD} (\texttt{CAlC})); see also \Cref{fig:features} in the appendix.

\subsection{Comparison with SMT and Quantifier Elimination Tools}

\begin{figure}[tb]
    \begin{subfigure}[t]{0.4\textwidth}
		\centering
        \includegraphics[scale=0.65]{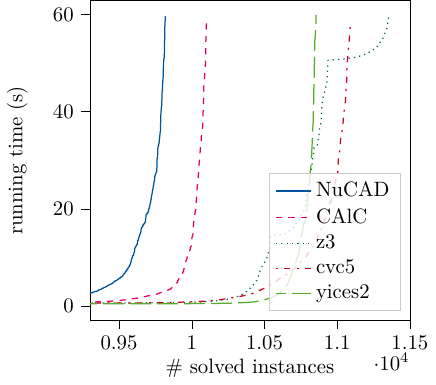}
        \caption{SMT-LIB QF\_NRA.}
        \label{fig:results_qfnra}
    \end{subfigure}\hfill
    \begin{subfigure}[t]{0.34\textwidth}
		\centering
        \includegraphics[scale=0.65]{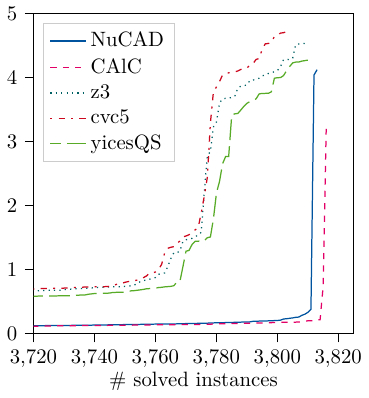}
        \caption{SMT-LIB NRA.}
        \label{fig:results_nra}
    \end{subfigure}\hfill
    \begin{subfigure}[t]{0.24\textwidth}
		\centering
        \includegraphics[scale=0.65]{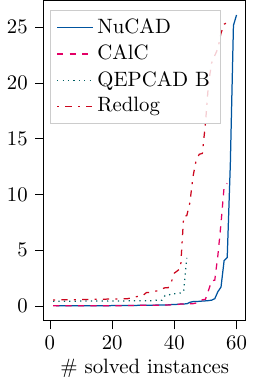}
        \caption{Wilson QE.}
        \label{fig:results_qe}
    \end{subfigure}
    \caption{Performance profiles of SMT solvers and QE tools.}
\end{figure}

We compare \texttt{NuCAD} and \texttt{CAlC} against the state-of-the-art SMT solvers \texttt{z3} 4.13.4, \texttt{cvc5} 1.2.0, and \texttt{yices2} 2.6.5 / \texttt{yicesQS}, as well as the open source quantifier elimination tools \texttt{QEPCAD B} (through \texttt{Tarski} 1.28) and \texttt{Redlog} 6658. The SMT results are depicted in \Cref{fig:results_qfnra} and \Cref{fig:results_nra}, the QE results in \Cref{fig:results_qe}. We omit the comparison against the \texttt{TIOpen-NuCAD} implementation from \cite{brown2015nucad,brown2017nucad} as it is incomplete and a fair comparison is not possible on the given benchmarks.

While \texttt{NuCAD} and \texttt{CAlC} are not competitive on QF\_NRA, both outperform the other solvers on NRA. On the QE benchmarks, \texttt{NuCAD} even solves the most instances. However, the NRA and QE benchmarks are quite small and not diverse, we thus cannot draw any robust conclusion. Unfortunately, there are no more accessible benchmark collections for these problems available currently. 

\section{Conclusion}
\label{sec:conclusion}

We reported on the first complete implementation of NuCAD. We simplified the algorithm for quantifier alternation by considering the quantifier structure during the computations, and applied further optimizations to reduce the computational effort. 

We evaluated the NuCAD algorithm against the related CAlC algorithm. We found that CAlC is more efficient on problems without quantifier elimination, as it produces fewer cells and avoids heavy computations with non-rational numbers more effectively. On the two benchmark sets with quantifier alternations, NuCAD was competitive or even more efficient; however, these sets are too small and not sufficiently diverse for drawing any reliable conclusion.
Combinations of the two algorithms could be explored in future to benefit from their individual strengths. In particular, NuCAD is better suited for parallelization: it computes a decomposition, and thus exactly describes the parts of the space which are yet unexplored. We thus can delegate their exploration to threads in a clean way. %

\printbibliography

\appendix

\section{Appendix}

\subsection{Improved Splitting}
\label{sec:nucad:improved}

The loop in Lines~\ref{alg:nucad:compute-nucad-full:splitting}-\ref{alg:nucad:compute-nucad-full:splittingend} of \Cref{alg:nucad:compute-nucad-full} splits the input cell $(\Isymb_{i+1},\ldots,\Isymb_{i'})$ into the generated cell $(\Isymb'_{i+1},\ldots,\Isymb'_{i'})$ and the neighbouring sections and sectors: if the generated cell is a sector, then we split the space below and above into a section and sector respectively, and explore them separately. However, in many cases, handling them separately causes unnecessary effort, as their truth value may be the same --- see e.g. the first and second child of $\texttt{T}$ in \Cref{fig:nucadds}.
In particular, exploring the sections is computationally heavy, as the points sampled from a section are likely non-rational.

\begin{algorithm}[b]
    \caption{Modification of \protect\FComputeNucadFull{}.}
    \label{alg:nucad:compute-nucad-full-mod}
    \uIf{$\Isymb'_j.l \neq \Isymb_j.l$}{
        {let} $? = {(}$ if $\Isymb_j.l$ is strict and $? = {[}$ otherwise \;
        \uIf{$\Isymb'_j.l$ is weak}{
            $Q := Q \cup \{ (\Isymb'_{i{+}1}, \ldots, \Isymb'_{j-1}, ?\Isymb_j.l,\Isymb'_j.l), \Isymb_{j+1}, \ldots, \Isymb_{i'}) \}$ \;
        }
        \uElse{
            $Q := Q \cup \{ (\Isymb'_{i{+}1}, \ldots, \Isymb'_{j-1}, ?\Isymb_j.l,\Isymb'_j.l], \Isymb_{j+1}, \ldots, \Isymb_{ i'}) \}$ \;
        }
    }
    \uIf{$\Isymb'_j.u \neq \Isymb_j.u$}{
        \tcp{analogous}
    }
\end{algorithm}

By using half-closed symbolic intervals, we can avoid such unnecessary splitting (or defer splitting in the corresponding recursive call for cases where the truth value indeed changes).
For doing so, we adapt \Cref{alg:nucad:compute-nucad-full,alg:nucad:compute-nucad-quantifier} as follows:
\begin{itemize}
    \item We allow closed bounds (e.g. $(l,u]$, $[l,u]$) for the symbolic intervals of the input cell and the generated cell.
    \item When adding the defining polynomials of the input cell to the polynomial set in \Cref{alg:nucad:compute-nucad-full:input} of \Cref{alg:nucad:compute-nucad-full}, we flag the ones that stem from closed bounds. For them, we do not require sign-invariance in the required cell, but only \emph{semi-sign-invariance} (it holds $p\leq 0$ or $p \geq 0$ on the required cell), which is introduced in \cite{nalbach2024mergingcells}.
    \item During the computation of the cell, we use the techniques described in \cite[Section 5]{nalbach2024mergingcells} which is able to derive half-closed cell bounds if possible, based on the mentioned flags and other properties.
    \item We replace the body of the \textbf{for}-loop in \Cref{alg:nucad:compute-nucad-full:splitting} of \Cref{alg:nucad:compute-nucad-full} by \Cref{alg:nucad:compute-nucad-full-mod}.
    \item Some cells may turn out empty when choosing a sample point; these cells need to be omitted.
\end{itemize}

\subsection{Encoding a NuCAD}
\label{sec:nucad:encoding}

For constructing a solution formula for quantifier elimination problems, we need to encode the NuCAD that describes the solution space. For doing so, we apply a post-processing to the NuCAD data structure --- similarly as done for the CAlC algorithm in \cite{nalbach2024calc}. The idea is --- given some node in the NuCAD data structure --- (1) to merge all neighbouring children which carry the same label ($\true$ or $\false$) and (2) to replace the subtree of a node where all of its leaves carry the same label by that label. 

\begin{example}
    Consider \Cref{fig:nucad}. The cell $[\iroot{x_1}{p_5}{1},\iroot{x_1}{p_5}{1}] \times (-\infty,\infty)$ is split into multiple cells which carry the same label, thus they can be merged during post-processing. Similarly, the cells after $(-\infty, \iroot{x_1}{p_8}{1}) \times [\iroot{x_2}{p_4}{1},\iroot{x_2}{p_4}{1}]$ and $(-\infty, \iroot{x_1}{p_8}{1}) \times [\iroot{x_2}{p_1}{1},\iroot{x_2}{p_1}{1}]$ can be merged respectively.\hfill\eoe
\end{example}

For obtaining the solution formula, we encode the parts of the tree that describe the cells labelled with $\true$. Without the preprocessing step, we would only need to encode the \emph{outermost} bounds on the variables (i.e. the bounds in the nodes closest to the leaves). However, after the preprocessing, some of these bounds are omitted; we thus need to include additional (non-outermost) bounds on variables. Further, in some cases, we would yield smaller solution formulas by taking the negation of the encoding of the cells labelled with $\false$. The work in \cite{nalbach2024calc} describes a technique that makes use of this observation and applies it also to subtrees of the NuCAD data structure.

\subsection{Evaluation}

\begin{table}[h]
    \caption{Features of solved instances, split by instances solved by both, none, or one solver exclusively. The features are the number of nodes (\# nodes) of the directed acyclic graph that encodes the input formula, and the maximal degree (max degree) of a single variable in a polynomial of the input formula. For each feature, we give the mean, median and maximal value over the corresponding instances.}
    \label{fig:features}
    \centering
    \begin{tabular}{l|r|rrr|rrr}
        \toprule
        & instances & \multicolumn{3}{c}{\# nodes} & \multicolumn{3}{c}{max degree} \\
        & \# solved & mean & median & max & mean & median & max \\
        \midrule
        both & 9028 & 430.08 & 65.00  & 6417 & 5.13 & 3  & 44 \\
        only \texttt{CAlC} & 232 & 833.12 & 306.50  & 4768 & 9.85 & 3  & 94 \\
        only \texttt{NuCAD} & 145 & 1565.12 & 2094.00 & 3703 & 3.80 & 2  & 16 \\
        none & 2749 & 13591.37 & 1427.00  & 664211 & 2.64 & 2  & 22 \\
        \bottomrule
    \end{tabular}
\end{table}

\end{document}